\documentclass[debug]{rmaa}
\usepackage{paralist}
\usepackage{psfrag,color}
\usepackage[latin1]{inputenc}

\def\msun{\mbox{M$_\odot$}}

\title{THE HELIUM AND HEAVY ELEMENTS ENRICHMENT OF THE GALACTIC DISK}

\author{
L. Carigi\altaffilmark{1,2}
and M. Peimbert\altaffilmark{1}
}

\altaffiltext{1}{Instituto de Astronom\'{\i}a, UNAM, M\'exico.}

\altaffiltext{2}{Centre for Astrophysics, UCLan, UK.}

\shortauthor{Carigi \& Peimbert}

\shorttitle{He enrichment of the Galactic ISM}

\fulladdresses{
\item
Both authors: Instituto de Astronom\'{\i}a, Universidad Nacional Aut\'onoma de M\'exico,
Apdo. Postal 70-264, M\'exico 04510 D.F., M\'exico
(carigi, peimbert@astroscu.unam.mx).
\item
Leticia Carigi:
Centre for Astrophysics, University of Central Lancashire,
Preston, Lancashire, PR1 2HE, United Kingdom
(lcarigi@uclan.ac.uk).
}

\listofauthors{L. Carigi \& M. Peimbert}
\indexauthor{Carigi, L.}
\indexauthor{Peimbert, M.}

\resumen{Presentamos modelos de evoluci\'on qu\'{\i}mica para el disco de nuestra galaxia. Tambi\'en presentamos una nueva determinaci\'on de $X$, $Y$, y $Z$ para M17, una regi\'on H~{\sc ii} de nuestra galaxia rica en elementos pesados. 
Comparamos nuestros modelos del disco gal\'actico con las abundancias de las regiones 
H~{\sc ii}. El valor predicho por nuestro modelo para $\Delta Y/\Delta O$ es muy similar al 
valor que obtenemos por medio de las observaciones de M17 y la abundancia primordial de helio, 
$Y_p$. A partir de M17 y $Y_p$ obtenemos que $\Delta Y/\Delta Z = 1.97 \pm 0.41$ resultado 
que concuerda con dos determinaciones de $\Delta Y/\Delta Z$, obtenidas a partir de observaciones 
de estrellas enanas K de la vecindad solar, que corresponden a $2.1 \pm 0.4$ y $2.1 \pm 0.9$ 
respectivamente. Nuestros modelos ajustan razonablemente bien el valor de O/H con el que se form\'o el Sol. }

\abstract{
We present chemical evolution models for the Galactic disk. We also present a new determination 
of $X$, $Y$, and $Z$ for M17 a Galactic 
metal-rich H~{\sc ii} region. We compare our models for the Galactic disk with the Galactic 
H~{\sc ii} regions abundances. The $\Delta Y/\Delta O$ ratio predicted from the galactic chemical evolution 
model is in very good agreement with the $\Delta Y/\Delta O$ value derived from M17 and the 
primordial helium abundance, $Y_p$, taking into account the presence of temperature variations in 
this H~{\sc ii} region. {From} the M17 observations we obtain that $\Delta Y/\Delta Z = 1.97 \pm 0.41$, 
in excellent agreement with two $\Delta Y/\Delta Z$ determinations derived from K dwarf 
stars of the solar vicinity that amount to $2.1 \pm 0.4$ and $2.1 \pm 0.9$ respectively. 
We also compare our models with the solar abundances. The solar and Orion nebula O/H values 
are in good agreement with our chemical evolution model. 
}

\addkeyword{galaxies: abundances}
\addkeyword{galaxies: evolution}
\addkeyword{H~II regions}
\addkeyword{ISM: abundances}
\addkeyword{ISM: individual: M17, Orion nebula}
\addkeyword{Sun: abundances}

\begin{document}

\maketitle

\section{Introduction}
\label{sec:intro}

The main purpose of this work is to study the evolution of the helium
abundance with respect to the heavy elements as a function of time and 
position in the Galactic disk. For this purpose we will use the Galactic
chemical evolution model by \citet{car05} that has been successful in 
explaining: the observed O/H and C/H abundance gradients in the interstellar medium, ISM, the 
present gaseous distribution in the Galactic disk,
the current star formation rate, the stellar mass as a function of the
Galactic radius, and most of chemical properties of the solar vicinity.

To have useful observational constraints we need accurate $X$, $Y$, and $Z$ 
determinations or at least $\Delta Y$/$\Delta Z$ determinations to compare
the models with observations. In this paper we recompute the $X$, $Y$, and $Z$
values for the H~{\sc ii} region M17, the best Galactic H~{\sc ii} region
for which it is possible to compute an accurate enough $Y$ value, for this purpose
we make use of the best observations available and the new He I atomic data
needed for the helium abundance determination. We also compare the chemical
evolution model with the $\Delta Y$/$\Delta Z$ determination derived from
K dwarf stars of the solar vicinity with metallicities similar or higher than
solar by \citet{jim03} and \citet{cas07}.

We also compare our model with the initial solar O/H value and with the Orion
O/H value. The initial solar O/H value is representative of the
ISM, 4.5 Gyr ago when the Sun was formed, and the Orion nebula O/H value is representative
of the present day ISM.

The model together with the primordial helium determination is also used to
provide an equation  between the $Y$ enrichment and the $O$ enrichment of the ISM. This 
equation can be used to provide the initial $Y$ values for those
stars for which we can derive their initial oxygen abundances. These initial $Y$ values
provide meaningful initial abundances for a set of stellar evolutionary models with different
heavy elements content.

We adopt the usual notation 
$X$, $Y$, and $Z$ to represent the hydrogen, helium and heavy elements abundances by mass, 
respectively. Based on our models we study the increase of helium,  $\Delta Y$, as a 
function of the increase of $C$, $O$, $Fe$, and $Z$ by mass. 

In Section~\ref{sec:models} we discuss the general properties of the
chemical evolution models, we discuss inflow models for the Galaxy
with two sets of stellar yields and present the chemical abundances for the disk 
at 5 galactocentric distances. For the two models discussed  we present the 
increase of helium by mass $\Delta Y$,
relative to the increase of carbon, oxygen, iron and heavy elements by mass,
$\Delta C$, $\Delta O$, $\Delta Fe$, and $\Delta Z$. We also discuss the evolution 
of the $Y$ and $O$ abundances for our models,  and present 
an equation that predicts for the Galaxy the $Y$ enrichment as a function of 
the $O$ enrichment of the ISM.  In Section~\ref{sec:M17} we 
present a new determination of the helium abundance for the metal rich
H~{\sc ii} region M17; this abundance is compared with our Galactic 
chemical evolution models.

In Section~\ref{sec:Orion} we compare abundances for the Orion nebula with those
of B stars of the Orion association. 
In Section~\ref{sec:solar O} we discuss the absolute calibration of the O/H 
ratio in the local ISM, this ratio is one of the most important observational 
constraints for Galactic chemical evolution models, the absolute calibration 
of the O/H ratio is obtained based on the Orion nebula and the solar 
abundances. In Section~\ref{sec:He} we compare the solar initial helium 
abundance, inferred from the standard solar models by \citet{bah06}, with 
our Galactic chemical evolution models considering the time since the Sun 
was formed and the presence of gravitational settling and diffusion. 
The conclusions are presented in Section~\ref{sec:conclusions}. 

Throughout
this paper we will use the primordial helium abundance by mass, $Y_p$, derived 
by \citet{pei07} based on direct helium abundance
determinations of metal poor extragalactic H~{\sc ii} regions that amounts to $0.2477 \pm 0.0029$.
This result is in excellent agreement with the $Y_p$ determination by \citet{dun08}
that amounts to $0.2484 \pm 0.0003$. This determination is based on the
$\Omega_bh^2$ value derived from WMAP observations, the assumption of standard big-bang
nucleosynthesis, and the neutron lifetime, $\tau_n$, of $885.7 \pm 0.8$ sec obtained 
by \citet{arz00}. Following \citet{mat05} the value of $Y_p$ derived from WMAP is revised downwards to 
$0.2468 \pm 0.0003$ by adopting the $\tau_n = 878.5 \pm 0.8$ sec derived by \citet{ser05}, and to
$0.2475 \pm 0.0006$ by adopting for $\tau_n$ the new world average that amounts to 
$881.9 \pm 1.6$ sec, average that includes the results by \citet{arz00} and \citet{ser05}.

\section{Chemical Evolution Models}\label{sec:models}

\subsection{Model Parameters}

\begin{figure}[!t]
\includegraphics[width=\columnwidth]{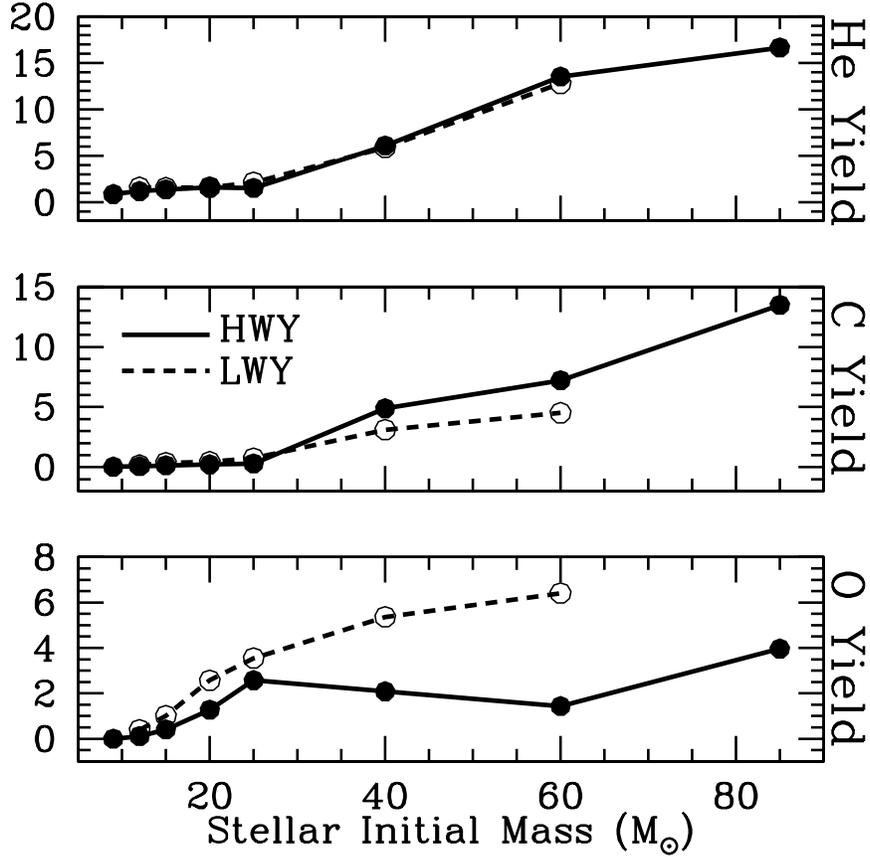}
\caption{
Newly formed mass of a given element by massive stars, in \msun, ejected to 
the ISM. The initial heavy elements of the stars amount to $Z=0.02$.
Continuous lines: high wind yields by \citet{mae92},
dashed lines: low wind yields by \citet{hir05}.
}
\label{fig:yields}
\end{figure}

We present chemical evolution models for the Galactic disk using the CHEMO code
\citep{car94} that considers the lifetime of each star until it leaves the main sequence. 
The models have been built to reproduce 
the present gas mass distribution and
the present-day O/H values for each galactocentric distance, 
the values were obtained from observations of H~{\sc ii} regions in the galaxy \citep{est05}.
Specifically, the characteristics of the models are:

i) An inside-outside scenario with primordial infalls but without any type of outflows.
The infall rate as a function of time and galactocentric distance $r$ is given by
$INFALL(r,t) =
A(r) e^{-t/\tau_{\rm halo}} + B(r) e^{-(t-1 Gyr)/\tau_{\rm disk}}$,
where the formation timescales are $\tau_{\rm halo}=0.5$ Gyr and
$\tau_{\rm disk}=6 + (r/r_\odot - 1)8$ Gyr.
We assume the location of the solar vicinity is $r_\odot=8$ kpc.
The constants $A(r)$ and $B(r)$ are chosen to match, first,
the present-day mass density of the halo and disk components
in the solar vicinity, 10 and 40 \msun $pc^{-2}$, respectively,
and second, to reproduce the radial profile total mass in the Galaxy,
$Mtot(r)= 50 e^{-(r-8)/3.5}$ \citep{fen03}.

ii)  13 Gyr as the age of the models, the time elapsed since the
beginning of the formation of the Galaxy.

iii) The Initial Mass Function (IMF) proposed by \citet{ktg93},
in the mass interval given by 0.01 $< m/\msun < M_{up}$,
with $M_{up}=$ 80 and 60 \msun.
This  IMF is a three power-law approximation, given by
IMF $\propto m^{-\alpha}$ with
$\alpha=-1.3$ for 0.01 - 0.5 \msun,
$\alpha=-2.2$ for 0.5 - 1.0 \msun,
and
$\alpha=-2.7$ for 0.5 - $M_{up}$.

Note that, our models were computed assuming an IMF with $M_{low} = 0.01$  \msun.
Kroupa et al. (1993) truncated their IMF at 0.08 \msun because they considered only stars,
but in our work we are assuming a non negligible amount of substellar objects
(0.01 $< m/\msun < 0.08$).
In models with $M_{up}$ = 60 \msun, the mass of objects with $ m< 0.08 \msun$
is ~12 \% of the total $M_{stars}$, that includes stars and remnants;
this percentage is practically independent of $M_{up}$.
Even at present, the fraction of mass in substellar objects is unknown and we consider
that our predicted percentage might be realistic.

Due to the uncertainties in the current $M_{gas}(r)$, $M_{stars}(r)$, and SFR$(r)$ values
we cannot discriminate between chemical evolution models assuming $M_{low} = 0.01$ \msun and $M_{low} = 0.08$ \msun.
The first ones predict smaller fractions of massive stars and LIMS per single stellar
generation. Therefore $M_{low} = 0.01$ \msun models with identical galaxy formation scenario, 
galactic age, and stellar yields to those of $M_{low} = 0.08$ \msun models, require a higher 
SFR to match the present-day O/H($r$) values. The $M_{low} = 0.01$ \msun model with a more 
efficient SFR predicts a lower $M_{gas}$ and similar chemical abundances. Such model with 
higher SFR and lower $M_{gas}$ is also able to reproduce the observational constraints \citep{car96}.

iv) A star formation rate that depends on time and galactocentric distance,
that varies from almost constant and low (at large $r$ values) to bursting and high
(at short $r$ values), this SFR has been represented by the following relation
$ SFR(r,t) = \nu  M^{1.4}_{gas}(r,t) \ (M_{gas}+M_{stars})^{0.4}(r,t)$,
in order to reproduce the current O/H gradient and the gas mass distribution of the Galactic disk \citep{car96},
where $\nu$ is a constant in time and space that is chosen in order to reproduce
the present-day radial distribution of the gas surface mass density.
A $\nu$ value  of 0.016 is required when the high-wind yields
and $M_{up}= 80$ \msun  are adopted, the best model of \citet{car05}, while
$\nu$ values  of 0.015 and 0.010 are required when the low-wind yields
with $M_{up}= 60$ and $M_{up}= 80$ \msun  are adopted, respectively.

v) Two sets of stellar yields. Since the main difference between these sets is the assumed mass-loss rate
due to stellar winds by massive stars with  $Z = 0.02$, we will call them high-wind yields (HWY) and low-wind yields (LWY),
see Figure 1. 

The HWY set is the one considered in the best model (model 1)
of \citet{car05}. The HWY set includes:
A) For massive stars (MS), those with $8 < m/\msun < 80$, 
the following yields by:
a) \citet{chi02} for  $Z = 0.00$;
b) \citet{mey02}
for  $Z = 10^{-5}$ and $Z = 0.004$;
c) \citet{mae92} for $Z = 0.02$ (high mass-loss rate yields presented in his Table 6);
d) \citet{woo95} only for the Fe yields
(Models B, for 12 to 30 \msun; Models C, for 35 to 40 \msun;
while for $m > 40 $ \msun, we extrapolated the $m = 40$ \msun Fe yields).
B) For  low and intermediate mass stars (LIMS), those with $0.8 \leq m/\msun \leq 8$,
we have used the yields by \citet{mar96,mar98} and \citet{por98} 
from $Z=0.004$ to $Z=0.02$.
C) For Type Ia SNe we have used the yields by \citet{thi93}.
We have assumed also that
5 \% of the stars with initial masses between 3 and 16 \msun \ are  binary systems which
explode as SNIa.

In the LWY set we have updated the yields of massive stars
only for $Z \sim 0$ and $Z = 0.02$ assuming the yields by \citet{hir07} and \citet{hir05} 
respectively. The rest of the stellar yields are those included in the high wind set.

The main differences between the LWY set and the HWY set are due to the contribution of
massive stars at $Z=0.02$. Therefore in Figure 1 we compare the He, C, and O yields for $Z=0.02$.
The main difference between the HWY and the LWY models is due to the stellar yields
assumed for massive stars
at high $Z$.
The HWY assume a relatively high mass-loss rate for massive stars with $Z=0.02$
\citep[yields by] []{mae92},
while the LWY assume a relatively low mass-loss rate for massive stars with $Z=0.02$ \citep[yields by][]{hir05}.
These difference between a high and a low mass-loss rate produces opposite differences in the C and O yields (see Figure 1),
the reasons are the following: a) a high mass-loss rate produces a high loss of C and consequently a
high C yield, and b) since C is needed to produce O, the high loss of C reduces the O yield.

Since the solar vicinity and the Galactic disk
contain stars and H~{\sc ii} regions of a broad range of metallicities,
our galaxy is a proper laboratory  
to study the $\Delta Y/\Delta X_i$ behavior at high $Z$ values
and to try to observationally test the predictions of the HWY models and the LWY models.

\begin{figure}[!t]
\includegraphics[width=\columnwidth]{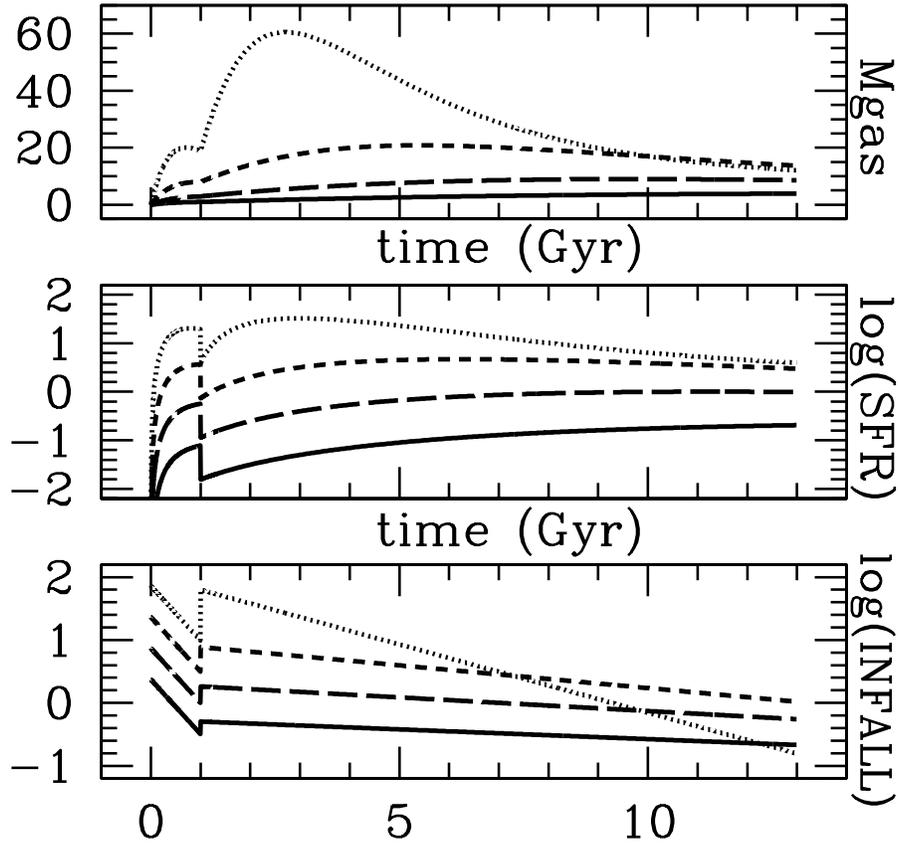}
\caption{
Evolution of some common properties for all models of the Galactic disk:
gas mass surface density ($M_\odot pc^{-2}$),
star formation rate and infall rates ($M_\odot pc^{-2} Gyr^{-1}$),
at four galactocentric distances, 16, 12, 8, and 4 kpc
(continuous, long-dashed, short-dashed, and dotted lines, respectively).
}
\label{fig:sfr}
\end{figure}

\subsection{Results}

The models presented in this paper reproduce the present stellar
and gas mass distributions in the Galactic disk,
the current star formation rate as a function of the Galactic radius,
the O/H gradient evolution inferred from PNe, the SN rates, 
the distribution of G-dwarf stars as a function of [Fe/H],
the infall rate, and the evolution of [X$_i$/Fe] vs [Fe/H].
See \citet{all98}, \citet{car94}, \citet{car96}, \citet{car00}.

In Fig. 2 we present some of those model properties,
like gas mass surface density, star formation rate, and infall rate
for four galactocentric distances: 16, 12, 8, and 4 kpc.
Infall rate follows the inside-outside scenario,
the inner parts of the Galaxy are formed faster
than the outer parts.
Since gas mass comes from infall, mainly, $Mgas$
reflects the inside-outside scenario and
the SFR shows similar behaviour as the gas mass,
due to the SFR is proportional to the gas mass.

\begin{figure}[!t]
\includegraphics[width=\columnwidth]{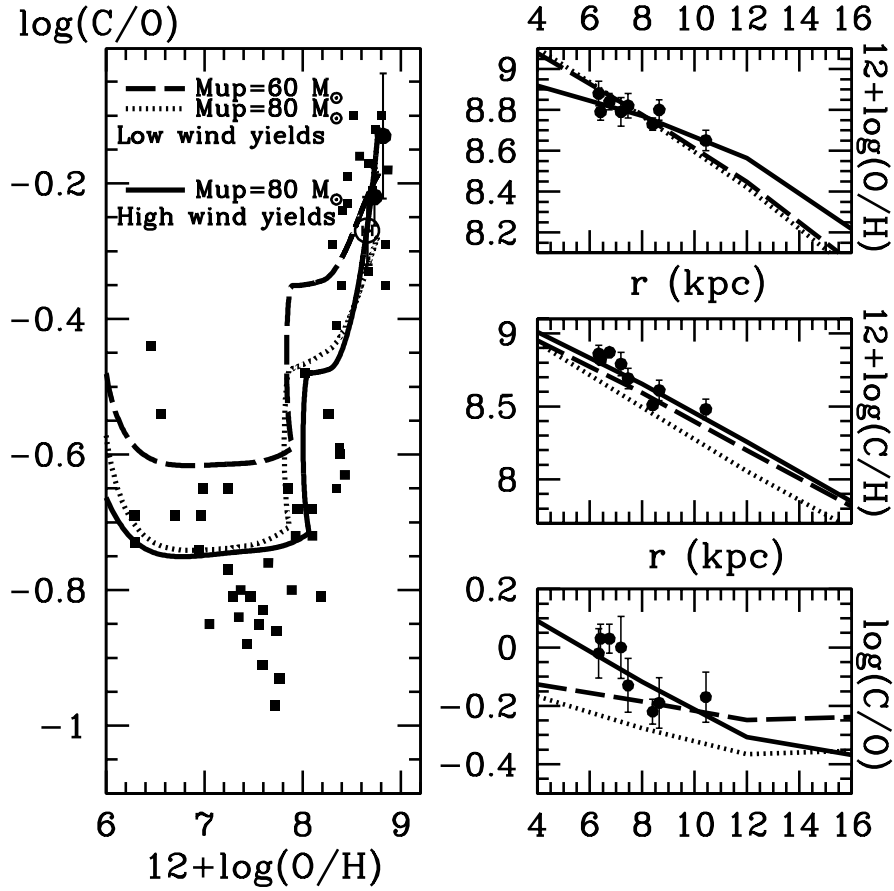}
\caption{Chemical evolution models for the Galactic disk and
the solar vicinity ($r=8$ kpc).
The left panel shows the C/O evolution in the ISM of the solar vicinity with O/H.
The right panels show the present-day ISM abundance ratios as a function of galactocentric distance.
Continuous lines: high wind yields with $M_{up} =80$ \msun  by \citet{car05} ,
dashed and dotted lines: low wind yields with  $M_{up} = 60$ and 80 \msun, respectively (this paper).
{\it Filled circles}: H~{\sc ii} regions, gas plus dust values; the gaseous values from \citet{est05}
 have been corrected for the dust fraction.
{\it Filled squares}: Dwarf stars from \citet{ake04}.
{\it Open circle}: Solar values from \citet{asp05}.
}
\label{fig:gradients}
\end{figure}

In Figure 3 we show O and C gradients in the Galactic disk and the evolution of the C/O-O/H relation
in the solar vicinity predicted by three models that combine different yields and IMF $M_{up}$ values.
Models that assume LWY fail to reproduce the C/O gradient and the C/O values for halo stars or disk stars. 
The LWY model with $M_{up} = 60$ \msun reproduces poorly the C/O gradient and the C/O values in disk stars,
and predicts C/O values for halo stars higher than observed.
The LWY model with $M_{up} = 80$ \msun does not reproduce at all the C/O gradient, 
matches partially the C/O values of disk stars, and explains the
observed C/O values for halo stars.
The HWY model with $M_{up} = 80$ \msun reproduces successfully the C/O Galactic gradient 
and the C/O values in the solar vicinity.

\begin{table}[!t]\centering
\setlength{\tabnotewidth}{0.5\columnwidth}
  \tablecols{6}
\setlength{\tabcolsep}{0.5\tabcolsep}
\caption{Present day values from the Galactic disk models }
\label{tta:disk}
\begin{tabular}{lccccc}
    \toprule
{Galactocentric distance} &
{$O (t_{final}) (10^{-3}$)}  &
{$\Delta Y/\Delta C$} &
{$\Delta Y/\Delta O$} &
{$\Delta Y/\Delta Fe$} &
{$\Delta Y/\Delta Z$} \\
\midrule
\multicolumn{6}{c}{High wind yields and $M_{up}=80$ \msun}\\
 4   & 8.89 & 6.38 & 5.91 & 14.11 & 1.85 \\
 8   & 6.68 & 7.01 & 4.01 & 18.36 & 1.67 \\
12   & 4.26 & 8.89 & 3.29 & 21.60 & 1.62 \\
16   & 1.95 & 10.05 & 3.23 & 25.28 & 1.65 \\
\multicolumn{6}{c}{Low wind yields and $M_{up}=60$ \msun}\\
 4   & 12.81 & 7.13 & 4.06 & 14.13 & 1.67 \\
 8   & 6.78 & 7.73 & 3.79 & 18.36 & 1.67 \\
12   & 3.28 & 9.44 & 3.99 & 21.52 & 1.78 \\
16   & 1.34 & 10.18 & 4.39 & 26.11 & 1.90 \\
\bottomrule
\end{tabular}
\end{table}

Based on Figure 3 we conclude that the LWY
model with $M_{up}$ = 60 \msun reproduces the main behavior of C/O vs O/H
in the solar vicinity  but
cannot reproduce the C/O Galactic gradient.
On the other hand, the HWY model reproduces very well the C/O vs O/H relation
in the solar vicinity and the C/O Galactic gradient. Since the LWY model with $M_{up}$ = 80 \msun
produces the poorest fit to the C/O and C/H observed values it will not be considered further.

\citet{gib06} considering the O yields by \citet{arn91}, that are lower  than
those by \citet{mae92} for MS with $m<30$ \msun,
reproduce the [O/Mg] values present in the Galactic Bulge.
With the same yields \citet{gib97} explains the
[O/Fe] values in the intracluster medium and
predicts a small increase in the C/O evolution at late times
for a massive elliptical galaxy. By adopting the \citet{arn91} yields in our model
we may obtain flatter C/O gradients and might not be able to reach
the C/O values observed in the H~{\sc ii} regions and dwarf stars
of the solar vicinity. The yields by \citet{arn91} 
do not consider stellar winds.
\citet{chi05} using LWY studied the C and O evolution in the solar vicinity
and the Galactic disk, 
they reproduce also the C/O vs O/H behavior in the solar vicinity, 
but they predict  flatter C/O gradients than the observed ones
and they cannot match the high C/O values shown by the metal rich star
in the solar neighborhood.

Recently, \citet{mcw07} suggested that the
strong metallicity-dependent yields 
for massive stars by \citet{mae92}, can explain the
O/Mg vs O-Mg/H and the O/Fe vs Fe/H relations in the Galactic bulge and in the solar vicinity,
in agreement with our results that favor the HWY model over the LWY model.
The HWY model includes low O yields at high $Z$ and therefore
explains: a) the small O increase in the solar vicinity
from the time the Sun was formed until the present, and b) the flattening of the O gradient
in the direction of the Galactic center. 
Massive stars with high $Z$ values have strong winds, lose a considerable amount of C
and produce high C yields.
With this C lost, the stars keep a small amount of C needed to produce O
and consequently their O yields are low.
Therefore the C yields for massive stars are important at high metallicities,
while their O yields are more important at low metallicities, as has been 
shown previously by \citet{ake04} and \citet{car05}.
By adding our previous results \citep{ake04,car05} to those of \citet{mcw07} and of this paper
we insist that the stellar winds with a high mass loss rate are essential
to reproduce the high C/O values observed in the disk stars of the solar vicinity.

In the upper panel of Figure 4 we present the evolution of the model
that assumes HWY and $M_{up} = 80$ \msun
at different galactocentric radii (4, 8, 12, and 16 kpc)
that correspond to different final metallicities ($Z=$ 0.028, 0.016, 0.009, and  0.004, respectively).
The $\Delta Y/\Delta O$ increase at $O > 4 \times 10^{-3}$ present in Figure 4 is due to the lower O yields
for massive stars with $Z=0.02$.
In the upper half of Table 1 we show the present-day $O$ values and the $\Delta Y/\Delta X_i$ values 
for each galactocentric radius.
We note that $\Delta Y/\Delta Z$ increases slightly with $Z$ for large $r$ values or low $O$ values, 
while $\Delta Y/\Delta Z$ increases significantly with $Z$ for short $r$ values or high $O$ values.

\begin{figure}[!t]
\includegraphics[width=\columnwidth]{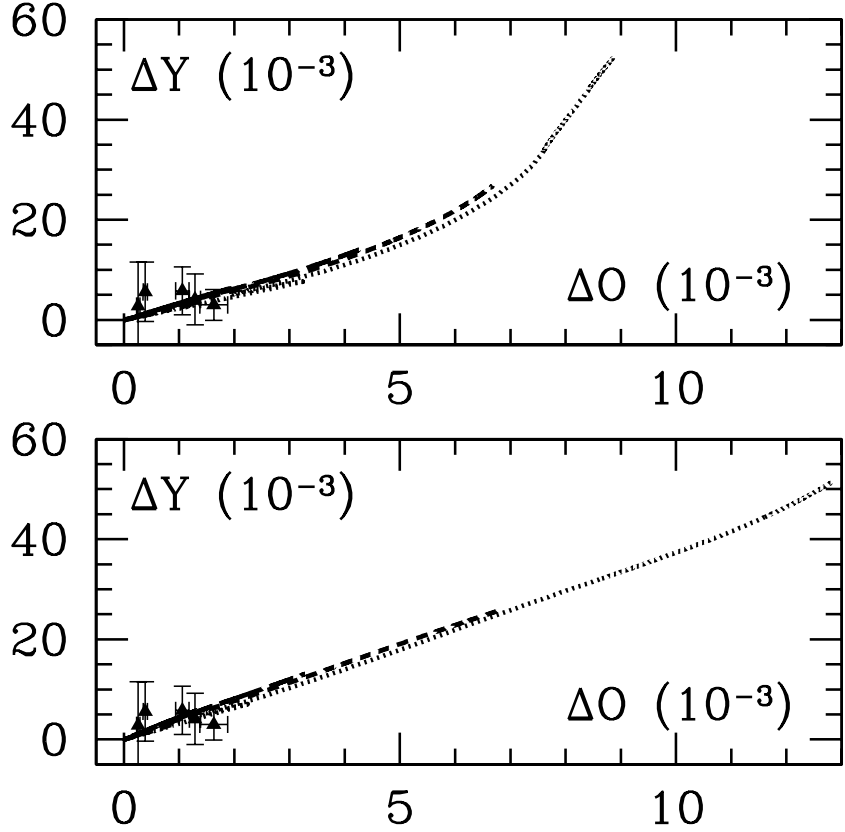}
\caption{Evolution of Helium vs Oxygen  for the Galactic disk at four
galactocentric distances, 16, 12, 8, and 4 kpc
(continuous, long-dashed, short-dashed, and dotted lines, respectively).
Upper panel:
The chemical evolution model assumes $M_{up}$ = 80 \msun and high wind yields.
Lower panel:
The chemical evolution model assumes $M_{up}$ = 60 \msun and low wind yields.
See Table 1.
Low metallicity H~{\sc ii} regions from \citet{pei07} ({\it filled triangles})
and adopting $Y_p$ = 0.2477.
}
\label{fig:disk6080}
\end{figure}

In the lower panel of Figure 4 we show the results for the model with LWY  and $M_{up} = 60$ \msun.
This model does not predict an increasing  $\Delta Y/\Delta O$ value with increasing $O$ for high $O$ 
values. 
In the lower half of Table 1 we show the present-day $O$ values and the $\Delta Y/\Delta X_i$ values 
for each galactocentric radius.
With the LWY model we find  higher final $O$ values for lower $r$ values because the O yields are higher than
than those of the HWY model at high $Z$.
Even if the HWY and LWY yields are identical for low $Z$,
we get lower final $O$ values for higher $r$ values for the LWY model because it does not
include stars with $m > 60$ \msun.
The increase or decrease of $O$ is reflected on the
$\Delta Y/\Delta O$ and $\Delta Y/\Delta Z$ values because the final $Y$ values are nearly
independent of the models.

The helium to oxygen mass ratio, $\Delta Y/\Delta O $, is an important 
constraint in the study of the chemical evolution of galaxies. 
We have studied the variation of $\Delta Y/\Delta O $ as $O$ increases
in the HWY and LWY evolution models (see Figure 4). 
For the HWY model, the model that fits the C/O gradient, 
we have found the following relations between $Y$ and $O$:

\begin{equation}
Y = Y_p + \Delta Y = Y_p + (3.3 \pm 0.7) O,  
\end{equation}
for $O < 4.3\times10^{-3}$, and 
\begin{equation}
Y = Y_p + (3.3 \pm 0.7) O + (0.016 \pm 0.003)(O/4.3 \times10^{-3} - 1)^2,
\end{equation}
for $ 4.3\times 10^{-3} < O < 9\times10^{-3}$.

For the LWY model with $M_{up}$ = 60 \msun we have found the following relation
between $Y$ and $O$:

\begin{equation}
Y = Y_p + (4.0 \pm 0.7) O,  
\end{equation}
for $0 < O < 11\times10^{-3}$. 

\citet{jim03} from a set of isochrones and observations of nearby 
K dwarf stars found that $\Delta Y/\Delta Z = 2.1 \pm 0.4$. \citet{cas07} found also that 
$\Delta Y/\Delta Z = 2.1 \pm 0.9$ from the newly computed set of Padova isochrones and observations of nearby 
K dwarf stars. These observational results are in very good agreement
with the models presented in Table 1.

Assuming yields with low mass-loss rate due to stellar winds \citep[similar to that considered by][]{hir07}, and yields without mass loss due to stellar winds by \citet{woo95}, \citet{chi03}
find for the solar vicinity $\Delta Y/\Delta Z \sim2.4$ and 1.5, respectively.

Since stellar winds change significantly the $C$ and $O$ yields,
but not the $Y$ yields,
we are interested to quantify the evolution of the  $Y$ contribution
due to massive stars and due to  low and intermediate mass stars
at different metallicities.
For that reason we show in Figure 5  
the accumulative percentage of $Y$ for four galactocentric distances due to MS and LIMS
obtained from the HWY model, our successful model for Galactic disk. 

\begin{figure}[!t]
\includegraphics[width=\columnwidth]{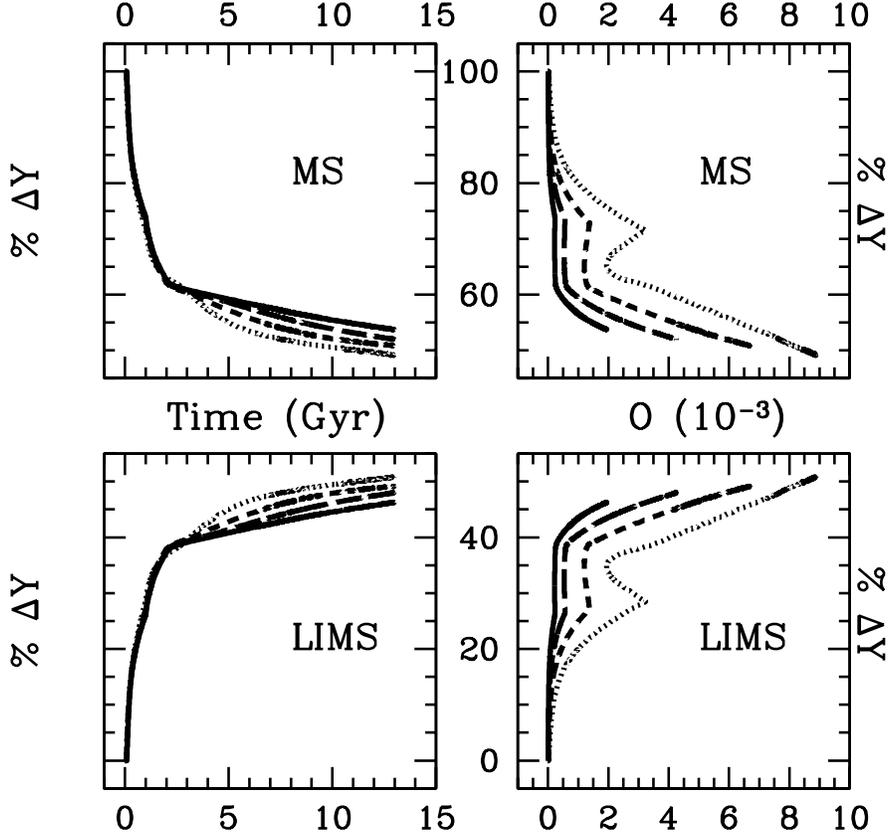}
\caption{
Cumulative percentage of He as a function of time and oxygen 
in the ISM, produced and ejected
by massive stars (MS) and low and intermediate mass stars (LIMS)
at four galactocentric distances on the Galactic disk (lines as Figure 2).
The model  assumes $M_{up}$ = 80 \msun and high mass loss due to stellar winds, the HWY model.
}
\label{fig:cumulative}
\end{figure}

The fraction of helium in the ISM due to MS and and to LIMS 
depends strongly on time, 
but not on galactocentric radius or $Z$.
At present about half of the $\Delta Y$ in the ISM  has been
produced by MS and half by LIMS.

The strong dependency on time of the $\Delta Y$ contribution is due to the lifetime of the stars.
The $\Delta Y$ contribution of LIMS decreases less than 5 \% from 4 to 16 kpc due  mainly to
the star formation history.
In the inside-outside scenario the SFR at 4 kpc is more intense and it
is also higher at earlier times than at larger distances
(see the middle panel in Figure 2), therefore the big number of LIMS formed 
in the first Gyrs at 4 kpc enrich the gas at later times.
This fact produces  the O dilution that can be seen in the right hand side 
panels of Figure 5, just after the model at 4 kpc reaches the value of $O = 3\times 10^{-3}$ 
for the first time.

\section{The helium and oxygen abundances of M17, a high metallicity Galactic H~{\sc ii} region}\label{sec:M17}

We will compare the predictions of the HWY and the LWY models with
observations of $Y$ and $O$ in the Galactic disk. At present the best H~{\sc ii} region
in the Galaxy to derive the $Y$ and $O$ abundances is M17. The reason is
that the correction for the presence of neutral helium in the abundance 
determination is the smallest for the well observed Galactic H~{\sc ii}  regions.
This is due to the high ionization degree of M17 \citep{pei92,est99,gar07}.
Due to the large amount of neutral helium present in the other well observed Galactic
H~{\sc ii}  regions, the error in the $Y$ determination is at least two times larger
than the error for M 17, therefore the $Y$ determinations for the other Galactic H~{\sc ii} 
regions will not be considered in this paper.

To determine very accurate He/H values of a given H~{\sc ii} region
we need to consider its ionization structure.  
For objects of low degree of ionization it is necessary to consider
the presence of He$^0$ inside the H$^+$ zone, while for objects of
high degree of ionization it is necessary to consider the possible
presence of a He$^{++}$ zone inside the H$^+$ zone.  \citet{pei92}, hereafter PTR, found for M17 an upper limit of
$N$(He$^{++})/N$(H$^+$) of 8 $\times 10^{-5}$ a negligible amount; alternatively
they found differences with position of the $N$(He$^{+})/N$H($^+$) ratio
correlated with the sulphur ionization structure, result that implies
that M17 is ionization bounded and the presence of a small but non
negligible amount of He$^0$ inside the H$^+$ zone. Therefore for this
object the helium abundance is given by
\begin{equation}
{{N({\rm He})}\over{N({\rm H})}} = 
{{N({\rm He}^{0})+N({\rm He}^{+})}\over{N({\rm H}^+)}}.
\end{equation}

To minimize the effect of the correction for neutral helium we took into account
only regions M17-1, M17-2, and M17-3, from now on M17-123, that show the highest 
degree of ionization of the observed regions by PTR, \citet{est99}, and
\citet{gar07}, as well as the highest accuracy in the line intensity determinations 
by PTR. Following PTR 
we will assume that He is neutral in the regions where S is once ionized, 
that is
\begin{equation}
{{N({\rm He}^{0})}\over{N({\rm He})}} = {{N({\rm S}^{+})}\over{N({\rm S})}}, 
\end{equation}
therefore
\begin{equation}
{{N({\rm He})}\over{N({\rm H})}} = {\rm ICF(He)} \times 
{{N({\rm He}^{+})}\over{N({\rm H}^+)}} = 
\left[1 + {{N({\rm S}^+)}\over{N({\rm S})-N({\rm S}^{+})}}\right] \times 
{{N({\rm He}^{+})}\over{N({\rm H}^+)}}  .  
\end{equation}

\noindent To estimate the ICF(He) value we recomputed the 
$N$(S$^{+})$/$N$(S) ratios derived by PTR taking
into account that the [S~{\sc ii}] $\lambda\lambda$ 4069 + 4076 lines are blended
with the O~{\sc ii} $\lambda\lambda$ 4069 and 4076 lines, the correction diminishes
the [S~{\sc ii}] electron temperatures from about 12 000 K to about 7700 K 
\citep{gar07}, the lower temperatures increase the $N$ (S$^{+})$/$N$(S)
ratios, and the average $ICF({\rm He})$ for M17-123 amounts to 1.035.

To obtain the $N$(He$^+$)/$N$H($^+$) value for M17-123 we decided 
to recompute the determinations by
PTR based on their line intensities and the new helium 
recombination coefficients by \citet{por05}, with the interpolation 
formula provided by \citet{por07}. In addition we used the hydrogen 
recombination coefficients by \citet{sto95},
and the collisional contribution to the He~{\sc i} lines by
\citet{saw93} and \citet{kin95}. The optical depth effects in the
triplet lines were estimated from the computations by \citet{ben02}.
At the temperatures present in M17 the collisional excitation of the hydrogen
lines is negligible and was not taken into account.

To determine the $N$(He$^+$)/$N$(H$^+$) value we took into account the
following He~{\sc i} lines $\lambda\lambda$ 3889, 4026, 4471, 4922, 5876, 
6678, and 7065. We corrected the 4922 line intensity by considering that it 
was blended with the [Fe~{\sc iii}] 4924 line and that
the contribution of the Fe line amounted to 5\% of the total line intensity
\citep{est99,gar07}. The M17-123 line intensities adopted are presented
in Table 2.

\begin{table}[!t]\centering
\setlength{\tabnotewidth}{0.5\columnwidth}
  \tablecols{2}
\caption{He~{\sc{i}} line intensities relative to H$\beta$ for M17-123}
\label{tta:HeLines}
\begin{tabular}{lc}
    \toprule
{He~{\sc i} Line} &
{$I$} \\ 
\midrule
3889  & $0.1738 \pm 0.0086$  \\
4026  & $0.0240 \pm 0.0012$  \\
4471  & $0.0508 \pm 0.0013$  \\
4922  & $0.0127 \pm 0.0010$  \\
5876  & $0.1549 \pm 0.0038$  \\
6678  & $0.0395 \pm 0.0010$  \\
7065  & $0.0499 \pm 0.0025$  \\
7281  & $0.0064 \pm 0.0007$  \\
\bottomrule
\end{tabular}
\end{table}

We did not correct the H and He line intensities for underlying absorption, the reasons are
the following: a) the average observed equivalent width in emission of H($\beta$), $EW_{em }$(H$\beta$),  amounts to
668 \AA\ , b) the predicted $EW_{em }$(H$\beta$) for  $T_e$ = 7000 K amounts to about
2000 \AA\ \citep{all84}, therefore about 1/3 of the
continuum is due to the nebular contribution and 2/3 to the dust scattered light from OB stars,
consequently the underlying stellar absorption only  affects two thirds of the observed continuum,
d) considering the nebular contribution to the observations and based on the models by \citet{gon99,gon05} 
for a model with an age of 2 Myrs as well as the observations by \citet{leo98}
for $\lambda$ 7065, \citep[since $\lambda$ 7065 was not included in the models by][]{gon99,gon05}, we 
estimated that the $EW_{ab }$ of the $\lambda\lambda$ 3889, 4026, 4471, 4922, 5876, 
6678, and 7065 lines amount to 0.4, 0.4, 0.4, 0.1, 0.1, 0.2 \AA\ respectively, an almost negligible amount
considering the large $EW_{em }$ observed values \citep[see Table 4 in][]{pei92}, e) the weighted
increase in the helium line intensities amounts to about 0.7\%, again an almost negligible amount,
f) the Balmer lines also show underlying absorption and the average correction to the line
intensities amounts to about 0.5\%, again a negligible amount that cancels to a first approximation 
the underlying correction effect on the He/H line ratios.

\begin{table}[!t]\centering
\setlength{\tabnotewidth}{0.5\columnwidth}
  \tablecols{3}
\caption{Physical conditions and chemical abundances in M17}
\label{tta:M17}
\begin{tabular}{lcc}
\toprule
{Parameter} &
{$t^2 = 0.000$}  &
{$t^2 = 0.036 \pm 0.013$}
\\
\midrule
$T$[O~{\sc ii} + O~{\sc iii}]       & $8300\pm 200$                     & $8300 \pm 200$  \\
$n$                       & $691 \pm246$                      & $744 \pm247$  \\
$\tau_{3889}$             & $9.5 \pm0.8$                      & $11.0 \pm 0.9$ \\
$N$(He$^+$)/$N$(H$^+$)    & $0.1014  \pm 0.0014$              & $0.0982 \pm 0.0019$ \\
$ICF$(He)                 & $1.035   \pm 0.010$               & $1.035 \pm 0.010$  \\
$N$(He)/$N$(H)            & $0.1049  \pm 0.0017$              & $0.1016 \pm 0.0022$ \\
$Y$                       & $0.2926  \pm 0.0034$              & $0.2837 \pm 0.0044$  \\
$\Delta Y$                & $0.0403 \pm  0.0044$\tablenotemark{a}   & $0.0360 \pm 0.0053$\tablenotemark{b} \\
$O$                       & $0.00446 \pm 0.00045$             & $0.00811 \pm 0.00081$ \\
$\Delta Y/\Delta O$       & $9.04 \pm 1.35$\tablenotemark{a}  & $4.44 \pm 0.79$ \tablenotemark{b} \\
$Z              $         & $0.0101 \pm 0.0015$               & $0.0183 \pm 0.0027$ \\
$\Delta Y/\Delta Z$       & $4.00 \pm 0.75$\tablenotemark{a}  & $1.97 \pm 0.41$ \tablenotemark{b} \\
\bottomrule
\tabnotetext{a}{Where we adopted $Y_p = 0.2523 \pm 0.0027$ for $t^2=0.000$ from \citet{pei07}.}
\tabnotetext{b}{Where we adopted $Y_p = 0.2477 \pm 0.0029$ for $t^2 \neq 0.000$ from \citet{pei07}.}
\end{tabular}
\end{table}

To determine the helium physical conditions of the nebula
simultaneously with the $N$(He$^+$)/$N$(H$^+$) value we used the maximum
likelihood implementation presented by \citet{pea02}. This implementation requires as inputs: a) the oxygen 
temperatures, $T$[O~{\sc iii}] and $T$[O~{\sc ii}], and the oxygen ionization degree
that provide us with the following restriction

\begin{equation}
T[{\rm O~{\scriptstyle{II}}} + {\rm O~{\scriptstyle{III}}}] = \frac{{N({\rm O}^{+})} {T[\rm O~{\scriptstyle{II}}]} +       
{{N({\rm O}^{++})}T[\rm O~{\scriptstyle{III}}]}} {{N({\rm O}^{+})}  + {N({\rm O}^{++})}},
\end{equation}

\noindent 
and b) a large set of helium to hydrogen line intensity ratios. In addition an estimate of the electron density, $n$, in the region where the He lines originate is not required but it is useful. This implementation
can determine the conditions of the H~{\sc ii} regions either with the restriction of uniform 
temperature, or relaxing this restriction.

{From} PTR we adopted $T$[O~{\sc iii}] = $8200 \pm 200$ K. {From} the $I$(3727/7325)
ratios for M17-123 by PTR and for M17-3 by \citet{est99}, after correcting the
$\lambda$ 7325~\AA\ lines for the recombination contribution \citep{liu00}, we
obtained $8100 \pm 1300$ K and  $9900 \pm 1300$ K respectively, therefore we
adopted for $T$[O~{\sc ii}] a value of $9000 \pm 1000$ K. {From} the $T$[O~{\sc iii}]
and $T$[O~{\sc ii}] values and the observations by PTR we find that $N$(O$^+$)/$N$(H$^+$) = 0.12
and $N$(O$^{++}$)/$N$(H$^+$) = 0.88. Finally from the previous results and equation (7) 
we obtained that $T$[O~{\sc ii} + O~{\sc iii}] = $8300 \pm 200$ K.

To estimate the electron density we used three determinations the $n$[S~{\sc ii}]
for M17-123 by PTR that amounts to $ 720 \pm 250$ cm$^{-3}$, the $n$[O~{\sc ii}]
for M17-3 by \citet{est99} that amounts to  $790 \pm 250$ cm$^{-3}$, and the
$n$[Cl~{\sc iii}] for M17-123 that we estimated from the observed $I$(5518)/$I$(5538)
ratios by PTR and the atomic physics parameters by \citet{kee00} that amounts to
$650 \pm 450$ cm$^{-3}$. {From} the average of these three determinations we adopted
a value of $n$ = $740 \pm 250$ cm$^{-3}$ for M17-123.

\begin{table}[!t]\centering
\setlength{\tabnotewidth}{0.5\columnwidth}
  \tablecols{6}
\setlength{\tabcolsep}{0.5\tabcolsep}
\caption{Present day values from the Galactic disk models for $r=6.75$
kpc
}
\label{tta:rm17}
\begin{tabular}{lccccc}
    \toprule
{Model} &
{$O (t_{final}) (10^{-3}$)}  &
{$\Delta Y/\Delta C$} &
{$\Delta Y/\Delta O$} &
{$\Delta Y/\Delta Fe$} &
{$\Delta Y/\Delta Z$} \\
\midrule
HWY $M_{up}=80$ \msun & 7.16 & 6.54 & 4.47 & 17.04 & 1.70 \\
LWY $M_{up}=60$ \msun & 8.28 & 7.26 & 3.73 & 17.04 & 1.62 \\
\bottomrule
\end{tabular}
\end{table}

By using as inputs for the maximum likelihood method $T$[O~{\sc ii} + O~{\sc iii}] = $8300 \pm 200$ K,
$n$ = $740 \pm 250$ cm$^{-3}$, and the helium line intensities presented in Table 2 we
obtain for M17-123 the $n$, $\tau_{3889}$, and  $N$(He$^+$)/$N$(H$^+$) values
presented in Table 3. The results for M17-123 are presented in Table 3 for $t^2 = 0.000$ 
(constant temperature over the observed volume) and for $t^2 \neq 0.000$ 
(the temperature variations method). Without 
the restriction of uniform temperature the maximum likelihood of the temperature 
fluctuation parameter amounts to $t^2 = 0.036 \pm 0.013$. This $t^2$ value is in good
agreement with those for M17 derived by PTR, \citet{est99}, and \citet{gar07}  that
are in the 0.033 to 0.045 range; these values were determined with two different
methods: a) combining the
temperature derived from the ratio of the Balmer
continuum to the Balmer line intensities with the temperature derived from
$I$(4363)/$I$(5007) [O~{\sc iii}] ratio,
and b) combining the O~{\sc ii} recombination line intensities with the $\lambda$ 5007 [O~{\sc iii}]
line intensities.

{From} the mean values of $N$(He$^+$)/$N$(H$^+$) given in Table 3 and the ICF(He) given by
equation (6)  we obtain $N$(He)/$N$(H) ratios for M17-123 of 0.1049 and 0.1016 for $t^2 = 0.000$ and 
$t^2 = 0.036$ respectively. These values are similar to but more precise than those derived 
by PTR for M17-123, that amount to 0.106 and 0.100 for  $t^2 = 0.000$ and $t^2 = 0.040$ respectively. 

\begin{figure}[!t]
\includegraphics[width=\columnwidth]{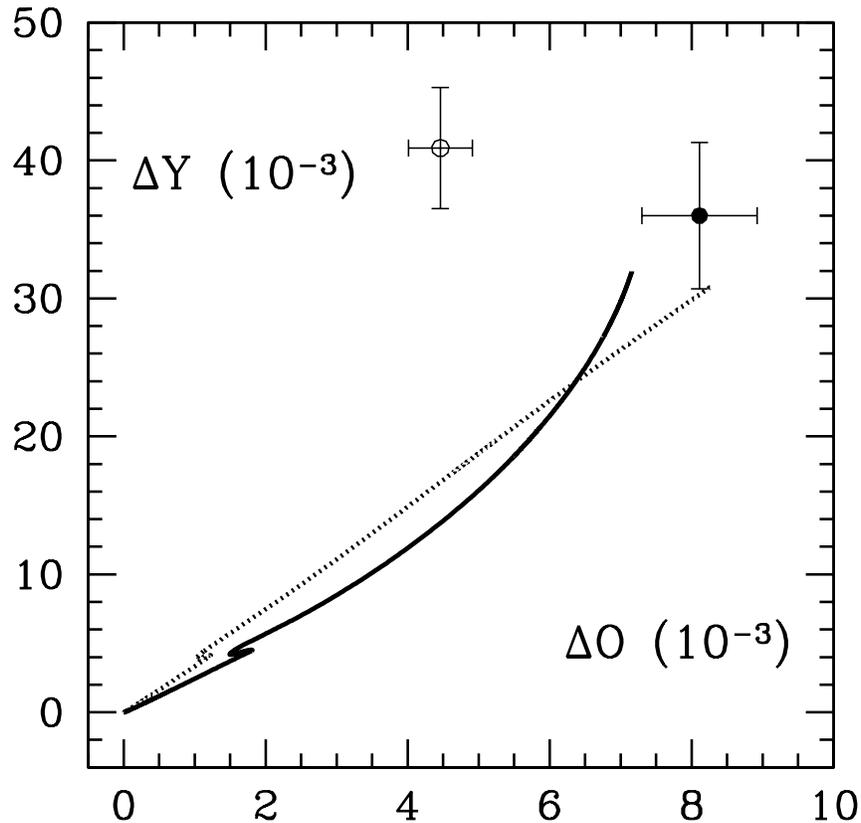}
\caption{Predicted evolution of He vs Oxygen for $r=6.75$ kpc.
Continuous line: model assumes $M_{up}$ = 80 \msun and high wind yield.
Dotted line: model assumes $M_{up}$ = 60 \msun and low wind yield.
M17 H~{\sc ii} region at $r=6.75$ kpc ({\it filled circle}, $t^2 \neq 0.000$)
 ({\it open circle}, $t^2 = 0.00$)}
\label{fig:diskm17}
\end{figure}

We obtained the $\Delta Y/\Delta O $ and the $\Delta Y/\Delta Z $
values presented in Table 3 based on the $Y_p$ determinations by \citet{pei07}. 
The $O$ abundance presented in Table 3 includes both the gaseous and the dust
contribution and corresponds to the average of the values derived by PTR, \citet{est99},
and \citet{gar07}, these three values are in excellent agreement.

M17 is located at a galactocentric distance of 6.75 kpc, under the assumption
that the Sun is located at a galactocentric distance  of 8 kpc \citep{dia02}.
In Figure 6 we have plotted 
the $\Delta Y/\Delta O $ value for $t^2 = 0.036$, from this figure it can be noted 
that this value is in good agreement, at about the one $\sigma$ level, with the Galactic chemical evolution
model based on the  HWY set,  and that the $O$ value corresponds to
the prediction by the models for a galactocentric distance of 6.75 kpc (see also
Table 4); alternatively the M17 $\Delta Y/\Delta O $ value for $t^2 = 0.000$ is 
considerably higher, by more than 3$\sigma$, than the value predicted by the HWY model. Similarly in Figure 6 we 
compare the M17 results with the LWY model, again the values for $t^2 = 0.036$ are in good agreement
with the Galactic chemical evolution model while the values for $t^2 = 0.000$ are not. Furthermore from
Table 4 it is also found that the $t^2 = 0.036$ value for $\Delta Y/\Delta Z$
is within one $\sigma$ of the models predictions, while the $t^2 = 0.000$ value for $\Delta Y/\Delta Z$
is about 3$\sigma$ away form the models predictions.

With the present accuracy of the $\Delta Y/\Delta O $ and $\Delta Y/\Delta Z$ determinations 
in Galactic H~{\sc ii} regions it is not possible to distinguish between the HWY and the LWY models.

\section{Comparison of stellar and nebular abundances in Orion}\label{sec:Orion}

To test the nebular abundances derived with different $t^2$ values for the Orion nebula we decided
to compare them with those derived for B star abundances of the Orion association. \citet{cun06} obtained
12 + log O/H = $ 8.70 \pm 0.09$ from 11 B stars, while \citet{lan08} obtained 12 + log Ar/H = $6.66 \pm 0.06$
from 10 B stars. These results are in excellent agreement with the nebular abundances derived by 
\citet{est04} for $t^2 \neq 0.00$ that amount to $8.73 \pm 0.03$ and $6.62 \pm 0.05$ for O and Ar respectively. Alternatively the values derived for $t^2 = 0.00$ amount to $8.59 \pm 0.03$ and $6.50 \pm 0.05$ for O and Ar
respectively, values that are about 1$\sigma$ and 3$\sigma$ smaller than those derived from B stars.
\section{Absolute calibration of the O abundance in the solar vicinity}\label{sec:solar O}

The predicted abundances by chemical evolution models are often compared with
stars and with H~{\sc ii}  regions abundances. The most popular comparisons are made
with the solar and the Orion nebula abundances. The comparisons among the
models, the Sun, and the Orion nebula are based on the absolute abundances,
therefore it is necessary to estimate not only the statistical errors but also the systematic
ones in the observational determinations.

We will start by considering the solar O/H abundance. What we want from the
Sun is the O/H value when it was formed, the so called initial value, and to keep in
mind that it is representative of the ISM 4.5 Gyr ago when the Sun
was formed. We have also to consider that the photospheric and the interior solar 
abundances might be different due to diffusion and gravitational settling.

In Table 5 we present the most popular O/H photospheric determinations of the last
fifty years, the quoted values and the errors are the original ones published by the authors. 
By looking at the differences among the different determinations it is clear that for
many determinations probably the errors represent only the statistical
errors and that systematic errors have not been taken into account, in short that the total 
errors have been underestimated. The determinations,
by \citet{asp05} and by \citet{all07}, are qualitatively different
to the previous five because they are based on 3D models while the other five
are based on 1D models. The last two determinations included in Table 5, those by \citet{caf08}
and \citet{cen08}, indicate that the possibility of a further revision of the solar abundance is still open.

The abundances inferred from interior models of the Sun, that are based on stereosismological data, 
are in disagreement with the 3D photospheric models and predict heavy element abundances 
about 0.2 dex higher than the 3D photospheric ones \citep[e. g.][and references therein]{bas07}.

\begin{table}[!t]\centering
\setlength{\tabnotewidth}{0.5\columnwidth}
  \tablecols{5}
\caption{12 + log(O/H)}
\label{tta:sunorion}
\begin{tabular}{lc@{\hspace{48pt}}ccc}
\toprule
&&\multicolumn{2}{c}{Orion nebula \tabnotemark{b}}&\\
\cline{3-4}
{Solar photosphere \tabnotemark{a}} & 
{Year} &
{$t^2 \neq 0.00$} &
{$t^2=0.00$}& 
{Year}
\\
\midrule
8.96            & 1960$^{(1)}$ & 8.79 $\pm$ 0.12 & ...& 1969$^{(11)}$ \\            
8.77 $\pm$ 0.05 & 1968$^{(2)}$ & 8.75 $\pm$ 0.10 & 8.52 $\pm$ 0.10 & 1977$^{(12)}$ \\
8.84 $\pm$ 0.07 & 1976$^{(3)}$ & ...             & 8.49 $\pm$ 0.08 & 1992$^{(13)}$ \\
8.93 $\pm$ 0.035& 1989$^{(4)}$ & 8.72 $\pm$ 0.07 & 8.55 $\pm$ 0.07 & 1998$^{(14)}$ \\
8.83 $\pm$ 0.06 & 1998$^{(5)}$ & ...             & 8.51 $\pm$ 0.08 & 2000$^{(15)}$ \\
8.736$\pm$ 0.078& 2001$^{(6)}$ & ...             & 8.49 $\pm$ 0.06 & 2003$^{(16)}$ \\
8.66 $\pm$ 0.05 & 2005$^{(7)}$ & 8.73 $\pm$ 0.03 & 8.59 $\pm$ 0.03 & 2004$^{(17)}$ \\
8.65 $\pm$ 0.03 & 2007$^{(8)}$ \\
8.86 $\pm$ 0.07 & 2008$^{(9)}$ \\
8.76 $\pm$ 0.07 & 2008$^{(10)}$ \\            
\bottomrule
\tabnotetext{a}{
1 \citet{gol60}. 2 \citet{lam68}. 3 \citet{ros76}. 
4 \citet{and89}. 5 \citet{gre98}. 6 \citet{hol01}. 7 \citet{asp05}. 8 \citet{all07}.
9 \citet{cen08}. 10 \citet{caf08}.}
\tabnotetext{b}{
11 \citet{pei69}. 12 \citet{pei77}.
13 \citet{ost92}. 14 \citet{est98}, this value includes the fraction of O 
tied up in dust grains that amounts to 0.08 dex. 15 \citet{deh00}. 
16 \citet{pil03}. 17 \citet{est04}, this value includes the fraction of O 
tied up in dust grains that amounts to 0.08 dex.}
\end{tabular}
\end{table}

To compare with our models we will use as the low O/H value the 3D photospheric 
determination by \citet{asp05} and as the high
O/H value the 1D photospheric determination by \citet{gre98}, that is in good agreement  
with the helioseismic determination. The next step is to have  reliable stellar interior
models to determine the initial O/H value, for this purpose we will use the models
by \citet{bah06} presented in Table 6.

The models by  \citet{bah06} indicate that the photospheric values of $Z$ and $Y$ do not
represent the initial values due to diffusion and gravitational settling. By assuming that the $O/Z$
ratio is not affected by these processes the initial O/H values correspond to
12 + log O/H = 8.70  and to 8.89 for the AGS05 and the GS98 surface abundances 
respectively. To obtain the present day ISM value we have
to consider that the Sun was formed 4.5 Gyr ago and according to the Galactic chemical
evolution model by \citet{car05} the O/H ratio in the ISM has increased by 0.13 dex
since the Sun was formed. Therefore our determinations of the present day ISM 12 + O/H values based on the
AGS05 and the GS98 abundances amount to 8.83 and 9.02 respectively.

In Table 5 we also present the most popular O/H determinations for the Orion nebula 
including the errors presented in the original papers. The predictions from the solar abundances 
and the chemical evolution models have to be compared with the total abundances in the 
nebula that have to include gas and dust. With the exception of the determinations 
by \citet{est98,est04} that take into account the dust fraction all the other 
determinations only include the gaseous content. The other difference is that there 
are two possible sets of values: a) those that assume constant temperature over the observed value given
by  the 4363/5007 ratio of [O~{\sc iii}], the $t^2$ = 0.00 case, where $t^2$ is the mean
square temperature variation \citep{pei67}, or b) those based on the O~{\sc ii} recombination lines, that
are in agreement with those derived from the 5007/H$\beta$ ratio taking into account
the presence of temperature variations over the observed volume,  
and consequently that $t^2 \neq 0.00$.

\citet{est04} obtain for the Orion nebula that 12 + O/H = 8.73 for $t^2 \neq 0.00$, 
value that includes the dust correction. By taking into account the presence of the O/H Galactic
abundance gradient that amounts to $-$0.044 dex kpc$^{-1}$ \citep{est05} we obtain a value of
12 + O/H = 8.75 for the local ISM. This value is smaller than the values estimated for the local
ISM based on: the solar photospheric abundances by AGS05 and the GS98, the standard solar models by
\citet{bah06}, and the chemical evolution of the Galaxy that amount to 8.83 and 9.02 respectively.
Similarly  from the results by \citet{est04} for the Orion nebula for $t^2 = 0.00$ and the observed
Galactic gradient we obtain that 12 + O/H = 8.51 for the local ISM. {From} the previous discussion it follows that
the best agreement for the derived O/H ISM value is given by the $t^2 \neq 0.00$ result from Orion and
the AGS0 result from the Sun. From these two determinations we recommend for the present day local ISM the value
12 + log O/H = $8.79 \pm 0.08$.

{From} the previous discussion it follows that the best agreement between the solar
and the Orion nebula abundances is obtained for the high nebular abundances, that
are derived from the $t^2 \neq 0.00$ values and that include the fraction of atoms tied
up in dust grains, and the AGS05 solar value.

\begin{table}[!t]\centering
\setlength{\tabnotewidth}{0.5\columnwidth}
  \tablecols{3}
\caption{Standard solar models \tablenotemark{a}}
\label{tta:Sun}
\begin{tabular}{lcr}
\toprule
{Values} &
{GS98}  &
{AGS05} \\
\midrule
initial $X$  & 0.70866 & 0.72594 \\
initial $Y$  & 0.27250 & 0.26001 \\
initial $Z$  & 0.01884 & 0.01405 \\
initial $O$  & 0.00879 & 0.00582 \\
initial O/H  & 8.89    & 8.70    \\
initial {$\Delta Y/\Delta Z$}& 1.32 & 0.88 \\
initial {$\Delta Y/\Delta O$}& 2.82 & 2.11 \\
surface $X$  & 0.7410  & 0.7586  \\
surface $Y$ \tabnotemark{b} & $0.2420 \pm 0.0072$ & $0.2285 \pm 0.0067$ \\
surface $Z$  & 0.0170  & 0.0122 \\
surface O/H \tabnotemark{b} & $8.83  \pm 0.17$   & $8.66 \pm 0.17$  \\
\bottomrule
\tabnotetext{a}{Standard solar models by \citet{bah06}. The GS98 and AGS05 columns correspond to models with the heavy element abundances derived from photospheric observations by \citet{gre98} and by \citet{asp05} respectively.}
\tabnotetext{b}{The errors are the conservative
ones adopted by \citet{bah06}.}
\end{tabular}
\end{table}

\section{The solar helium abundance}\label{sec:He}

We want to compare also our Galactic chemical evolution model with the helium
abundance when the Sun was formed, the initial $Y$ value. \citet{bas04} based on 
seismic data have derived the $Y$ value in the solar convective envelope and 
amounts to $0.2485 \pm 0.0034$. To derive the initial value we need a model
of the solar interior that takes into account helium and heavy elements diffusion
and that agrees with the helium abundance of the envelope.

Again we have at our disposal the solar interior models by \citet{bah06} presented
in Table 6. The GS98 model agrees with the $Y$ value in the envelope derived by \citet{bas04}, but not with
the O/H value in the envelope derived by \citet{asp05}. On the other hand the AGS05 model agrees with
the O/H value in the envelope derived by \citet{asp05} but not with the $Y$ value derived by \citet{bas04}.
Based on the discussion of the previous section we conclude that the Orion nebula O/H value agrees
with the O/H value predicted by the Galactic chemical evolution model and the photospheric
value by \citet{asp05} and not with the photospheric value by \citet{gre98}. The discrepancy
between the photospheric abundances by \citet{asp05} and the $Y$ value derived from helioseimological
data is a very important open problem, an excellent review discussing this issue has been
presented by \citet{bas07}.

\citet{bah06} present the initial $Y$,
$Z$, and $O$ solar values for their standard solar models, see Table 6. 
By adopting the $Y_p$ value by \citet{pei07} for $t^2 \neq 0.00$  it is also
possible to obtain the $\Delta Y/\Delta Z$ and the $\Delta Y/\Delta O$ values for the GS98 and 
the AGS05 standard solar models. The
values so derived are in fair agreement with the predictions of the HWY and LWY Galactic chemical
evolution models. To try to make a more rigorous comparison between the solar interior and Galactic
chemical evolution models it is necessary to estimate the errors in the initial solar values 
by \citet{bah06}, a task which is beyond the scope of this paper.

\section{Conclusions}\label{sec:conclusions}

Based on the HWY model we find the following equation 
to estimate the initial helium abundance with which stars form in the Galactic disk

$$Y = Y_p + (3.3 \pm 0.7) O,$$  
for $O < 4.3\times10^{-3}$, and 

$$Y = Y_p + (3.3 \pm 0.7) O + (0.016 \pm 0.003)(O/4.3 \times10^{-3} - 1)^2,$$
for $ 4.3\times 10^{-3} < O < 9\times10^{-3}$.

The increase of {$\Delta Fe/\Delta Z$} has to be taken into account in order to
determine the $\Delta Y/\Delta Z $ value based on the [Fe/H] abundances of
stars in the solar vicinity.

High mass loss rates due to stellar winds should be adopted in the evolutionary stellar models
for massive stars of high metallicity, because
only the Galactic chemical evolution models with HWY can reproduce
simultaneously the O/H  and C/O Galactic gradients, the $C/O$ versus $O$ relation 
in the solar vicinity, and the $\Delta Y/\Delta O$ value in the inner Galactic disk.

Based on the O/H value of the Orion nebula and the solar photospheric value together 
with a chemical evolution model of the Galaxy we recommend for the present day local ISM  
a value of 12 + log O/H = $8.79 \pm 0.08$, where both
the gaseous and the dust components of O are taken into account.

By comparing the O/H value of the Orion nebula with the solar value we find
that the nebular ratio derived using O recombination lines, that is
equivalent to the use of the forbidden O lines under the adoption of a $t^2 \neq 0.00$, is in considerably better
agreement with the initial solar value than the Orion nebula value derived adopting $t^2 = 0.00$.

The stellar O/H and Ar/H abundance ratios derived by \citet{cun06} and \citet{lan08} for B 
stars of the Orion association are in excellent agreement with the nebular abundance ratios 
derived from the $t^2 \neq 0.00$ values for the Orion nebula.

The $\Delta Y/\Delta Z  = 1.97 \pm 0.41$ value derived from observations of M17 
for $t^2 = 0.036$ is in very good agreement with the $2.1 \pm 0.4$ and the $2.1 \pm 0.9$ values derived 
by \citet{jim03} and \citet{cas07} from K dwarf stars of the solar vicinity.
On the other hand the value $\Delta Y/\Delta Z  = 4.00 \pm 0.75$ derived from 
observations of M17 for $t^2 = 0.000$ is not. 

Both Galactic chemical evolution models with the
HWY set and the LWY set are in agreement with the observed $\Delta Y/\Delta Z$ for $t^2 = 0.036$
but not with the $\Delta Y/\Delta Z$ for $t^2 = 0.000$. Higher accuracy determinations
of $\Delta Y/\Delta Z$ for high metallicity objects are needed to discriminate
between the HWY model and the LWY model predictions.

\vskip 1cm
 
We are grateful to Brad Gibson and Antonio Peimbert for several fruitful discussions. We are also
grateful to the anonymous referee for some excellent suggestions. This work was 
partly supported by the CONACyT grants 46904 and 60354.

\end{document}